\documentclass[12pt,preprint]{aastex}
\slugcomment{Published in ApJ Letter 617 (2004) L9}

\shorttitle{Evolution of Close Galaxy Pairs in DEEP2}
\shortauthors{Lin et al.}

\begin{document}

\title{THE DEEP2 GALAXY REDSHIFT SURVEY: EVOLUTION OF CLOSE GALAXY PAIRS AND MAJOR-MERGER RATES UP TO z $\sim$ 1.2 \altaffilmark{1}}

\author{LIHWAI LIN \altaffilmark{2,3}, DAVID C. KOO \altaffilmark{3}, CHRISTOPHER N.
A. WILLMER \altaffilmark{3,4}, DAVID R. PATTON \altaffilmark{5},
CHRISTOPHER J. CONSELICE \altaffilmark{6}, RENBIN YAN
\altaffilmark{7}, ALISON L. COIL \altaffilmark{7}, MICHAEL C.
COOPER \altaffilmark{7}, MARC DAVIS \altaffilmark{7,8}, S. M.
FABER \altaffilmark{3}, BRIAN F. GERKE \altaffilmark{8}, PURAGRA
GUHATHAKURTA \altaffilmark{3}, AND JEFFREY A. NEWMAN
\altaffilmark{9}}

\altaffiltext{1}{ Some of the data presented herein were obtained
at the W. M. Keck Observatory, which is operated as a scientific
partnership among the California Institute of Technology, the
University of California, and the National Aeronautics and Space
Administration. The Observatory was made possible by the generous
financial support of the W. M. Keck Foundation.}
\altaffiltext{2}{Department of Physics, National Taiwan
University, No. 1, Sec. 4, Roosevelt Road, Taipei 106, Taiwan;
d90222005@ntu.edu.tw} \altaffiltext{3}{University of California
Observatories/Lick Observatory, Department of Astronomy and
Astrophysics, University of California at Santa Cruz, 1156 High
Street, Santa Cruz, CA 95064.} \altaffiltext{4}{On leave from
Observat\'{o}rio Nacional, Brazil.} \altaffiltext{5}{Department of
Physics, Trent University, 1600 West Bank Drive, Peterborough, ON
K9J 7B8, Canada.} \altaffiltext{6}{Department of Astronomy,
California Institute of Technology, Pasadena, CA 91125.}
\altaffiltext{7}{Department of Astronomy, University of California
at Berkeley, 601 Campbell Hall, Berkeley, CA 94720-3411.}
\altaffiltext{8}{Department of Physics, University of California
at Berkeley, 366 LeConte Hall, Berkeley, CA 94720-7300.}
\altaffiltext{9}{Hubble Fellow; Institute for Nuclear and Particle
Astrophysics, Lawrence Berkeley National Laboratory, Berkeley, CA
94720.}

\begin{abstract}
\hspace{3mm}We derive the close, kinematic pair fraction and
merger rate up to redshift $z \sim 1.2$ from the initial data of
the DEEP2 Redshift Survey. Assuming a mild luminosity evolution,
the number of companions per luminous galaxy is found to evolve as
$(1+z)^{m}$, with $m= 0.51\pm0.28$; assuming no evolution, $m =
1.60\pm0.29$. Our results imply that only $9\%$ of present-day
$L^{*}$ galaxies have undergone major mergers since $z\sim1.2$ and
that the average major merger rate is  about $4\times 10^{-4}$
$h^{3}$ Mpc$^{-3}$ Gyr$^{-1}$ for $z \sim 0.5 - 1.2$. Most
previous studies have yielded higher values.
\end{abstract}


\section{INTRODUCTION}
Galaxy interactions and mergers are an integral part of our
current paradigm of the hierarchical formation and evolution of
galaxies. Such processes are expected to affect the morphologies,
gas distributions, and stellar populations of galaxies (e.g.,
Mihos \& Henquist 1994, 1996; Dubinski, Mihos, \& Henquist 1996).
Although mergers are rare today (Patton et al. 2000, hereafter
P2000), cold dark matter $N$-body simulations show merger rates of
halos increasing with redshift as $(1+z)^{m}$, with $2.5\leq
m\leq3.5$ \citep{gov99,got01}.

Observations, however, have yielded results with $0\leq m\leq4$
\citep{zep89,bur94,car94,woo95,yee95,
pat97,neu97,le00,car00,pat02,bun04}. The diverse  results are
likely due to different pair criteria, observational techniques,
selection effects, and cosmic variance (Abraham 1999; P2000). To
identify close pairs, the most secure method is via spectroscopic
redshifts for both galaxies to find kinematic pairs \citep[
hereafter P2002]{pat02}. This Letter adopts the approach of P2002
and uses the early data of the DEEP2 (Deep Extragalactic
Evolutionary Probe 2) Redshift Survey (Davis et al. 2003; S. M.
Faber et al. 2004, in preparation) to derive the pair fraction and
merger rates up to $z=1.2$. In $\S$2, we describe the sample and
selection functions. In $\S$3, we use both the projected
separation and the relative velocity to select close galaxy pairs
and to determine their evolution. Major merger rates out to
$z\sim1.2$ are computed in $\S$4. The results are discussed in
$\S$5. Throughout this Letter, we adopt a cosmology of $H_{0}=70$
km s$^{-1}$ Mpc$^{-1}$, $\Omega_{m}=0.3$, and
$\Omega_{\Lambda}=0.7$.

\section{DATA AND SELECTION FUNCTIONS}
The DEEP2 Redshift Survey (DEEP2 for short) will measure redshifts
for $\sim50,000$ galaxies at $z\sim 1$ \citep{dav03} using DEIMOS
(DEep Imaging Multi-Object Spectrograph; Faber et al. 2003) on the
10 m Keck II telescope. The survey convers four fields, with Field
1 (Extended Groth Strip) being a strip of 0.25 $\times$ 2
deg$^{2}$ and Fields 2, 3, and 4 each being 0.5 $\times$ 2
deg$^{2}$. The photometry is based on $BRI$ images taken with the
12k $\times$ 8k camera on the Canada-France-Hawaii Telescope (see
Coil et al. 2004 for details). Galaxies are selected for
spectroscopy using a limit of $R_{AB}=24.1$ mag. Except in Field
1, a two-color cut is also applied to exclude galaxies with
redshifts $z < 0.75$. A 1200 line mm$^{-1}$ grating (R $\sim$
5000) is used with a spectral range of $6400 \-- 9000$ \AA, where
the [O II] $\lambda$3727 doublet would be visible at $z\sim0.7 \--
1.4$. The data are reduced using an IDL pipeline developed at
UC-Berkeley (J. A. Newman et al. 2004, in preparation). The
$K$-correction is derived from spectra of local galaxies (C. N. A.
Willmer et al. 2004, in preparation). The data used here are from
Fields 1 and 4, cover $\sim 0.4$ deg$^{2}$, and contain $\sim5000$
galaxies.

To measure  the spectroscopic selection functions  (see L. Lin et
al. 2004, in preparation, for details), we compared the sample
with successful redshifts to galaxies in the full photometric
catalog that satisfy  the limiting magnitude and two-color cuts.
Following analogous approaches by \citet{yee96} and P2002, we
calculate the spectroscopic weight $w$ of each galaxy as $1/S$,
where $S$ is the spectroscopic selection function derived from the
$R$ magnitude, $B-R$ and $R-I$ colors, $R-$band surface brightness
$\mu_{R}$, and local galaxy density of the galaxy itself. To
correct for bias due to slit collisions \citep{dav04}, we also
compute the angular weight, $w(\theta)$, as a function of angular
separation $\theta$ (P2002). For $\theta \leq 20"$, we find $0.95
\leq w(\theta) \leq1.5$, a result confirmed with the DEEP2 mock
catalog of \citet{yan04}.

\section{PAIR STATISTICS}
A major limitation on the direct measurement of merger fractions
for galaxies is the difficulty in identifying on-going mergers,
especially for distant galaxies. One alternative is to count only
the pairs with projected separations $\vartriangle$$r$ and
relative line-of-sight heliocentric velocities $\vartriangle$$v$
less than $r_{max}$ and $v_{max}$, respectively. A large fraction
of pairs with physical separations less than 20 $h^{-1}$ kpc and
velocity difference less than 500 km\ s$^{-1}$ appear to have
disturbed morphologies or signs of interactions, and these
galaxies are expected to merge within 0.5 Gyr (P2000). In our
work, close pairs are defined such that their projected
separations satisfy 10 $h^{-1}$ kpc $ \leq $ $\vartriangle r$
$\leq r_{max}$ and their rest-frame relative velocities
$\vartriangle$$v$ are less than 500 km\ s$^{-1}$. We adopt an
inner cutoff with a projected distance 10 $h^{-1}$ so as to avoid
the ambiguity between very close pairs and single galaxies with
multiple  star-forming knots. We choose values of $r_{max}$ = 30,
50, and 100 $h^{-1}$ kpc, where 30 $h^{-1}$ kpc is most likely to
include genuine merger pairs, while the two larger separations
provide larger samples and thus better statistics. To ensure the
selection of the same types of galaxies at different redshifts  in
the presence of  luminosity evolution, P2002 adopted a specific
range in evolution-corrected absolute magnitude $M_{B}^{e}$,
defined as $M_{B}+Qz$, where the evolution is parameterized as
$M(z)=M(z=0)-Qz$. Following P2002, we adopt $Q=1$ as a primary
model, and we also study the effect of different models on the
pair statistics. We restrict the analysis to galaxies with
luminosities $-21\leq M_{B}^{e}\leq -19$ for $z=0.45 - 1.2$. Since
there is only a 2 mag range in our sample, and assuming a constant
$M/L$ ratio, most observed pairs are thus major mergers, i.e.,
mass ratios between 1 : 1 and 6 : 1. Using data from the DEEP2
Fields 1 and 4  helps us reduce the effects of  cosmic variance.
Following P2000 and P2002, we compute the average number of
companions per galaxy
\begin{equation}
N_{c}=\frac{\sum^{N_{tot}}_{i=1}\sum_{j}w_{j}w(\theta)_{ij}}{N_{tot}},
\end{equation}
where $N_{tot}$ is the total number of galaxies within the chosen
magnitude range, $w_{j}$ is the spectroscopic weight for the
$j$th companion belonging to the $i$th galaxy, and
$w(\theta)_{ij}$ is the angular weight for each pair as described
in Section 2. Details of the weighting scheme can be found in
Section 5 of P2002 and references therein. This quantity $N_{c}$
is similar to the pair fraction when there are few triplets or
higher order $N$-tuples in the sample, which is the case here. In
this work, $N_{c}$ will sometimes simply be referred to as the
pair fraction.

 Figure 1 shows  $N_{c}$ versus redshift $z$ for a
sample with $v_{max} = 500$ km\ s$^{-1}$ and 10 $h^{-1}$ kpc
$\leq$
 $\vartriangle$$r \leq r_{max}$ with $r_{max}= $ 30, 50, and 100 $h^{-1}$ kpc, respectively. In the case of using
$r_{max}=$ 50 $h^{-1}$ kpc, we find 79 paired galaxies out of 2547
galaxies. The derived $N_{c}$ is $\sim8\%$ at $z\sim0.6$ and
increases to 10\% at $z\sim1.1$. Figure 1 also shows results from
the SSRS2 (Southern Sky Redshift Survey; P2000) at $z\sim 0.015$
and CNOC2 (Canadian Network for Observational Cosmology; P2002) at
$z\sim 0.3$, after corrections that adopt the same cosmology and
the same luminosity range.

The DEEP2 sample is $R$-band-selected and hence, for redshifts
greater than 0.8, is biased toward galaxies that are bright in the
rest-frame UV, i.e., against  faint red galaxies, especially at
redshifts $z
>1.0$.
To avoid this bias, we divide
our sample into blue and red galaxies using the color bimodality feature
and then repeat the
calculation of $N_{c}$ using $r_{max}=$ 50 $h^{-1}$ kpc for blue
galaxies at $z$ = 0.3 (CNOC2), 0.6, and 0.85 bins (DEEP2). When
using only blue galaxies, the values of the pair fraction
$N_{c}^{blue}$ decrease at $z$ = 0.3 and 0.85 but increase at $z$
= 0.6 compared to the results using the original sample. The
changes in the measured value of $N_{c}$ could be up to a factor
of 2,  depending on the field and redshift. To mitigate the
underestimation of the pair fraction in the redshift bin $1.0<z<1.2$,
we calculate a corrected $N_{c}$ at $z=1.1$ as
\begin{equation}
N_{c}^{cor}=\max[N_{c},2N_{c}^{blue}],
\end{equation}
shown as open triangles in Figure 1. Parameterizing the evolution
of $N_{c}$ as $Nc(0)(1+z)^{m}$, we find that the best fit of
$(N_{c}(0),m)$ is $(0.029\pm0.005,1.08\pm0.40)$ with
$\chi^{2}=0.27$ for $r_{max}=30$ $h^{-1}$ kpc,
$(0.068\pm0.008,0.51\pm0.28)$ with $\chi^{2}=0.89$ for
$r_{max}=50$ $h^{-1}$ kpc, and $(0.177\pm0.014,0.47\pm0.18)$ with
$\chi^{2}=1.2$ for $r_{max}=100$ $h^{-1}$ kpc. For the highest
redshift bin, $N_{c}^{cor}$ instead of $N_{c}$ is used for the
above fitting procedure. The best fits are shown as long-dashed
curves in Figure 1. For three different choices of $r_{max}$, we
find only a small increase of $N_{c}$ from redshifts $z=0$ to
$z=1.2$. In figure 1, we also list the upper limits of errors from
the cosmic variance for each redshift bin calculated by assuming a
spherical geometry of the survey volume. Incorporating the cosmic
variance into the fitting procedure slightly raises $m$ to
$0.62\pm0.49$ for $r_{max}=50$ $h^{-1}$ kpc.

The adopted choice of $Q = 1$ in the luminosity evolution model
sets the luminosity range of the sample at each redshift and thus
affects our results. To assess this effect, we repeat the pair
analysis for other choices of $Q$. The luminosity ranges are
chosen such that they are locked to $-22\leq M_{B}\leq -20$ at
$z=1$. Using $r_{max}=$ 50 $h^{-1}$ kpc and an upper redshift
limit of $z = 1$, we find that $m$ varies from $1.60\pm0.29$ for
$Q = 0$, $0.86\pm0.29$ for $Q = 0.5$, $0.41\pm0.30$ for $Q = 1$,
to $-0.24\pm0.35$ for $Q = 2$. In figure 2, we plot $N_{c}$ as a
function of absolute $B$-band magnitude ($M_{B}$) for three
redshift samples. A clear dependence of $N_{c}$ on $M_{B}$ is
evident. The trend that $m$ decreases with $Q$  can be understood
as the result of including fainter galaxies and thus having larger
$N_{c}$ at lower redshifts when adopting higher $Q$ values.
Nevertheless, even with the lowest choice of $Q = 0$ to maximize
$m$, models with $m>3.5$ are ruled out at a 3 $\sigma$ level of
confidence for all three choices of separation limits.

\section{MAJOR MERGER RATES}
The comoving merger rate, usually defined as the number of mergers per
unit time per comoving volume, can be
estimated as
\begin{equation}
N_{mg}=0.5n(z)N_{c}(z)C_{mg}T^{-1}_{mg},
\end{equation}
where $T_{mg}$ is the time-scale for physically associated pairs
to merge, $C_{mg}$ denotes the fraction of galaxies in close pairs
that will merge in $T_{mg}$, and $n(z)$ is the comoving number
density of galaxies. The factor 0.5 is to convert the number of
galaxies into the number of merger events. The merger rate
calculated above, however, is not suitable for comparison to the
merger rate derived from morphological approaches. The reason is
that while we have restricted the luminosity range of companions
to compute $N_{c}$, the morphological approach does not. To
correct for this restriction, equation (3) can be modified :
\begin{equation}
N_{mg}=(0.5+G)n(z)N_{c}(z)C_{mg}T^{-1}_{mg},
\end{equation}
where the added parameter $G$ accounts for the excess number of
companions failing to fall into our sample. Assuming that the
maximum mass ratio needed to yield significant morphology
distortions is 4 : 1, we calculate the value of $G$ as 1.24 for
$-21\leq M_{B}^{e}\leq -19$ using the local luminosity function
\citep{dri03}. $T_{mg}$ depends on the relative mass ratio,
dynamical orbit, and detailed structure of the two merging
galaxies, and thus it varies from case to case. Here we adopt
$T_{mg}$ = 0.5 Gyr, a value suggested by $N$-body simulations and
simplified models (Mihos 1995; P2000). The value $C_{mg}$ is
approximately 0.5 for close pairs with 5 $h^{-1}$ kpc $\leq$
$\vartriangle$$r$ $\leq 20$ $h^{-1}$ kpc and $v_{max} = 500$ km\
s$^{-1}$; this estimate is based on morphological studies of local
close pairs (P2000). Nevertheless, both $T_{mg}$ and $C_{mg}$
remain uncertain.

The derived merger rates are displayed in Figure 3. Here we have
applied the pair fraction evolution of $m=0.51$ derived for the
$r_{max}=50$ $h^{-1}$ kpc case to the $z\sim0$ (SSRS2) $N_{C}$
result for pairs with 5 $h^{-1}$ kpc $\leq$ $\vartriangle$$r$
$\leq 20$ $h^{-1}$ kpc. The parameter $n(z)$ is calculated as the
sum of the number of galaxies in the adopted magnitude range, each
weighted by its spectroscopic weight and divided by the comoving
volume occupied by the included sources. The errors shown for
DEEP2 measurements represent $40\%$ variations that are typical
for close pair counts in our sample, and they do not include the
uncertainties of $T_{mg}$ and $C_{mg}$. Also shown are the results
from \citet{con03}, who relied on morphologically identified
mergers for $M_{B}<-19$ using \emph{Hubble Space Telescope}
(\emph{HST}) data in the Hubble Deep Field-North (\emph{filled
triangles}). Clearly, the comoving merger rate in our data changes
little  with redshift. From $z\sim0.5$ to $z\sim1.2$, the average
is $N_{mg} \sim4\times 10^{-4}$ $h^{3}$ Mpc$^{-3}$ Gyr$^{-1}$ for
$-21\leq M_{B}^{e} \leq -19$. This value is about 1 order of
magnitude lower than the average merger rate for galaxies with
$M_{B} \leq -19$ derived by Conselice et al. (2003).

Finally, we calculate the merger remnant fraction, $f_{rem}$,
defined as the fraction of present-day galaxies that have
undergone major mergers (P2000). Adopting the merger fraction
$C_{mg}$ to be 0.5 and $T_{mg}$ to be 0.5 Gyr, we estimate, using
equation (32) in P2000, that about $9\%$ of present $L^{*}$
luminous galaxies have undergone major mergers since $z\sim1.2$.

\section{DISCUSSION}
Studies up to $z\sim 1$ using pair counts have found a wide range
in the evolution of merger rates. For example, based on roughly
300 Canada-France Redshift Survey galaxies measured with
\emph{HST}, Le F\`{e}vre et al. (2000) concluded that the pair
fraction evolves with $m=2.7\pm0.6$ while the fraction of merger
candidates evolves with $m=3.4\pm0.6$. In contrast, Carlberg et
al. (2000) measured the mean fractional pair luminosity from $z$ =
0.2 to 1 and found no evolution with $m~\sim0\pm1.4$, consistent
with our result. Since the pair fraction depends on adopted
luminosity limits, differences among studies may depend on the
choice of luminosity evolution models. \citet{le00}, e.g., applied
no corrections while \citet{car00} adopted the Q=1 model, as we do
here. The key advantages of our pair sample over previous surveys
include having measurements at $z > 1$, a larger sample, and more
restrictive pair criteria than those adopted by \citet{le00} and
\citet{car00}.

The discrepancy in the derived merger rates at $z\sim1$ between
our estimates and those from \citet{con03} is more difficult to
reconcile. Besides the choice of luminosity ranges and the
uncertainties in $C_{mg}$ and $T_{mg}$, other factors may explain
the discrepancy. First, cosmic variance can always come into play,
since most morphological studies have been forced to use the few
small fields covered deeply by \emph{HST} imaging, whereas
pair-count surveys cover much larger areas. Second, the two
approaches may sample interacting and merging  galaxies at
different stages, with morphological approaches identifying very
advanced mergers and merger remnants, while the pairs detect some
of the same systems before distortions are discernible. Matching
the derived merger rates from these two approaches requires
knowledge of the precise timescales of close pairs and of the
duration time for the appearance of distorted morphologies.
However, neither is well understood yet. Third, while our close
pairs identify only major mergers, morphological criteria may be
sensitive to minor mergers as well. Finally, both the pair-count
and morphology methods are subject to systematic uncertainties
that require detail modeling (Bell 2004). Given the current
discrepancy in results, it behooves us to study the connection
between kinematic pairs and morphologically disturbed galaxies at
various redshifts, both via observational approaches (P2000; J.
Lotz et al. 2004, in preparation) and through numerical
simulations (T. J. Cox et al. 2004, in preparation).

The merger rates from pair counts are also an excellent test for
those estimated from $N$-body simulations, since the former reveal
the behavior of luminous baryons while the latter reflect the
behavior of dark halos. Gottl\"{o}ber et al. (2001) defined merger
rates in $N$-body simulations as equivalent to the major merger
fraction per gigayear by tracing the formation and evolutionary
history of each halo. These fractions will differ from the pair
fractions roughly by a constant, and therefore their redshift
evolution is amenable to direct comparisons. They found merger
rates evolving as $(1+z)^{3}$ up to $z\sim2$. This theoretical
value of $m = 3$ is significantly higher than the $m \sim 0.5$ we
find from observed pair counts.  We should, however, be wary about
this comparison for the following reasons. First, definitions of
merger rates in observations and simulations may not be consistent
with each other, since the halos and visible galaxies span
different size scales and since $N$-body simulations also suffer
limitations in resolution. Second, the merger rates/fractions of
halos are likely to be a function of halo mass, which is suggested
by our finding that the galaxy pair fraction is a function of
luminosity. Pair count works using $K$-band-selected samples
(e.g., Bundy et al. 2004) can also provide another avenue for
testing the mass dependence of merger rates since the $K$-band
luminosity is more representative of any underlying stellar mass
and suffers less from evolution corrections. Although preliminary,
mock catalog simulations by E. Van Kampen et al. (2004, in
preparation) predict flat slopes for the pair-count fractions (see
Fig. 3 of Bell 2004), just as seen in our observations. We expect
a dramatic improvement in our understanding of merger histories
via pair counts after more realistic comparisons to simulations
are possible and especially after a 10-fold increase in sample
size when DEEP2 is complete in 2005.

\acknowledgments We thank the referee for helpful comments, J.
Primack, J. Lotz, and B. Allgood for useful discussions, and the
Keck staff for dedicated assistance. L. L. acknowledges support
from Taiwan via the COSPA project and NSC grant
NSC92-2112-M-002-021, D. R. P. from NSERC of Canada, and the DEEP2
team from NSF grants AST00-71048, AST00-71198, and KDI-9872979. We
close with thanks to the Hawaiian people for allowing us to use
their sacred mountain.

\clearpage

\begin{figure}[t]
\plotone{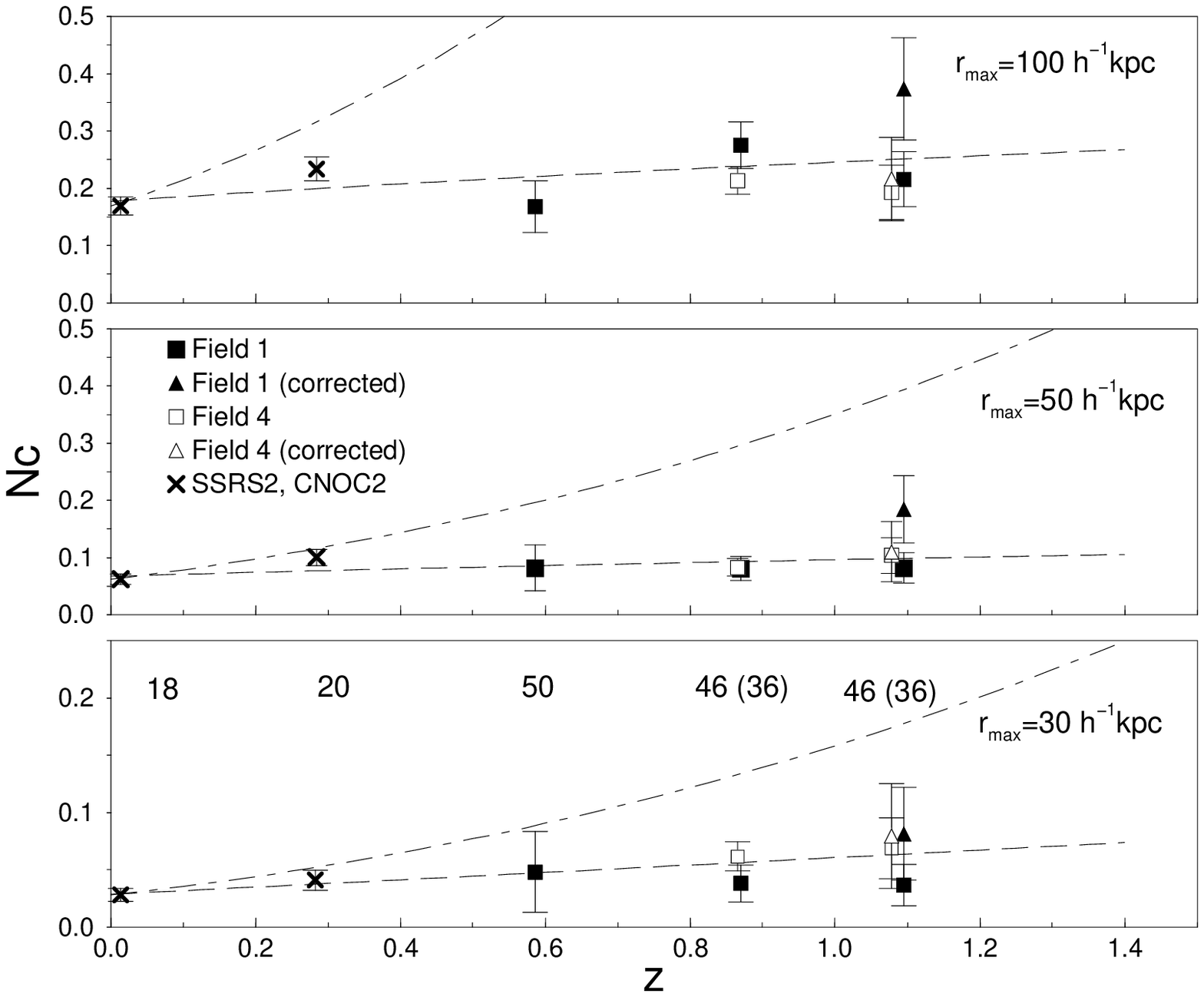} \caption{Pair fraction, $N_c$, as a function of
redshift $z$. The filled squares represent measurements from DEEP2
Field 1, and the open squares represent those from Field 4. The
crosses mark the results from SSRS2 and CNOC2. The open triangles
show the corrected pair fraction $N_{c}^{cor}$ for the highest
redshift bin. The long-dashed curves are best fits by the form
$(1+z)^{m}$, using DEEP2, SSRS2, and CNOC2 data (see text); the
dot-dashed curves represent evolution with $m=2.5$ for reference.
From the top to the bottom, $r_{max}=100$ $h^{-1}$ kpc,
$r_{max}=50$ $h^{-1}$ kpc, and $r_{max}=30$ $h^{-1}$ kpc. Note
that the bottom panel has a different vertical scale. The error
bars shown in the plot and used for fitting are calculated by
bootstrapping. The numbers appearing in the bottom panel indicate
the upper limits of errors (in units of percentage) from the
cosmic variance for each redshift bin. Those in the parentheses
are for open symbols. \label{fig1}}
\end{figure}

\clearpage

\begin{figure}[t]
\plotone{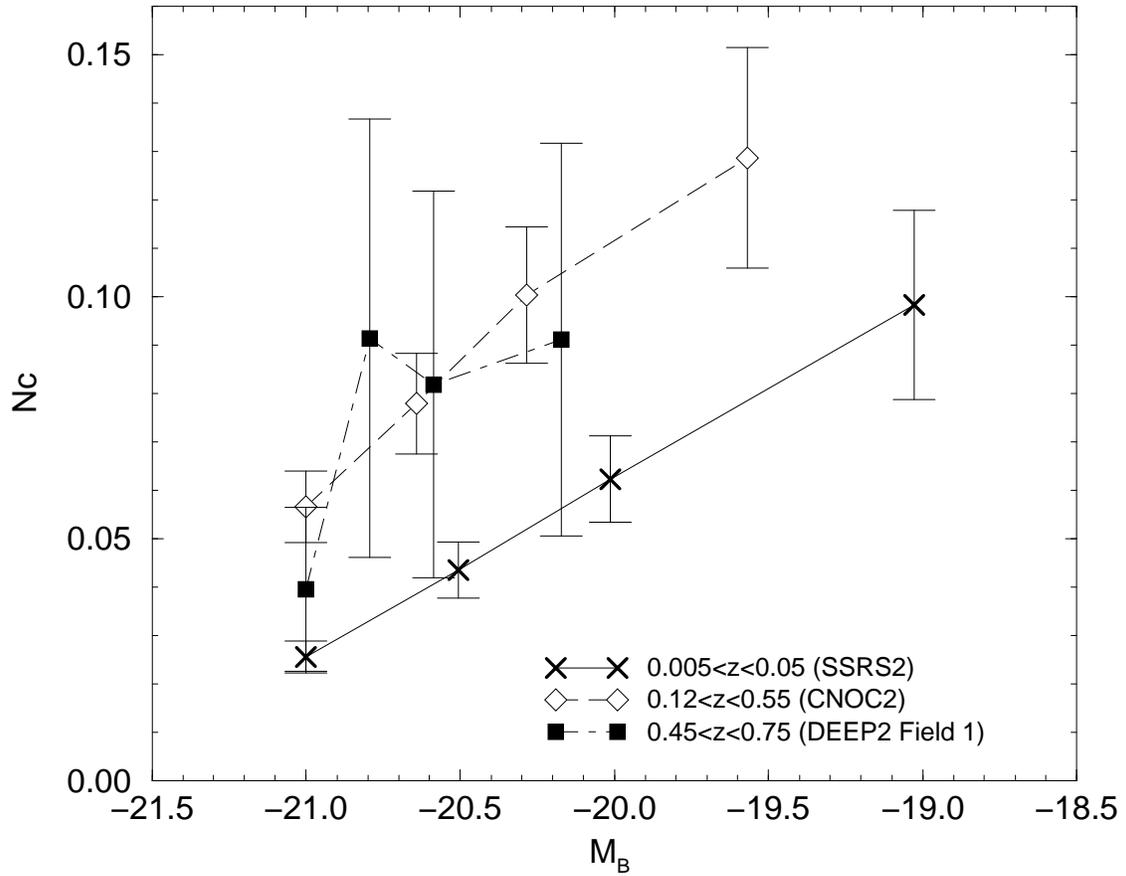} \caption{Pair fraction, $N_c$, as a function of
absolute $B$-band magnitude ($M_{B}$) using $r_{max}=50$ $h^{-1}$
kpc for different redshift samples as indicated. For each data
point, $N_c$ is calculated using galaxies with absolute $B$-band
magnitude between $M_{B}-1$ and $M_{B}+1$. \label{fig2}}
\end{figure}

\clearpage

\begin{figure}[t]
\plotone{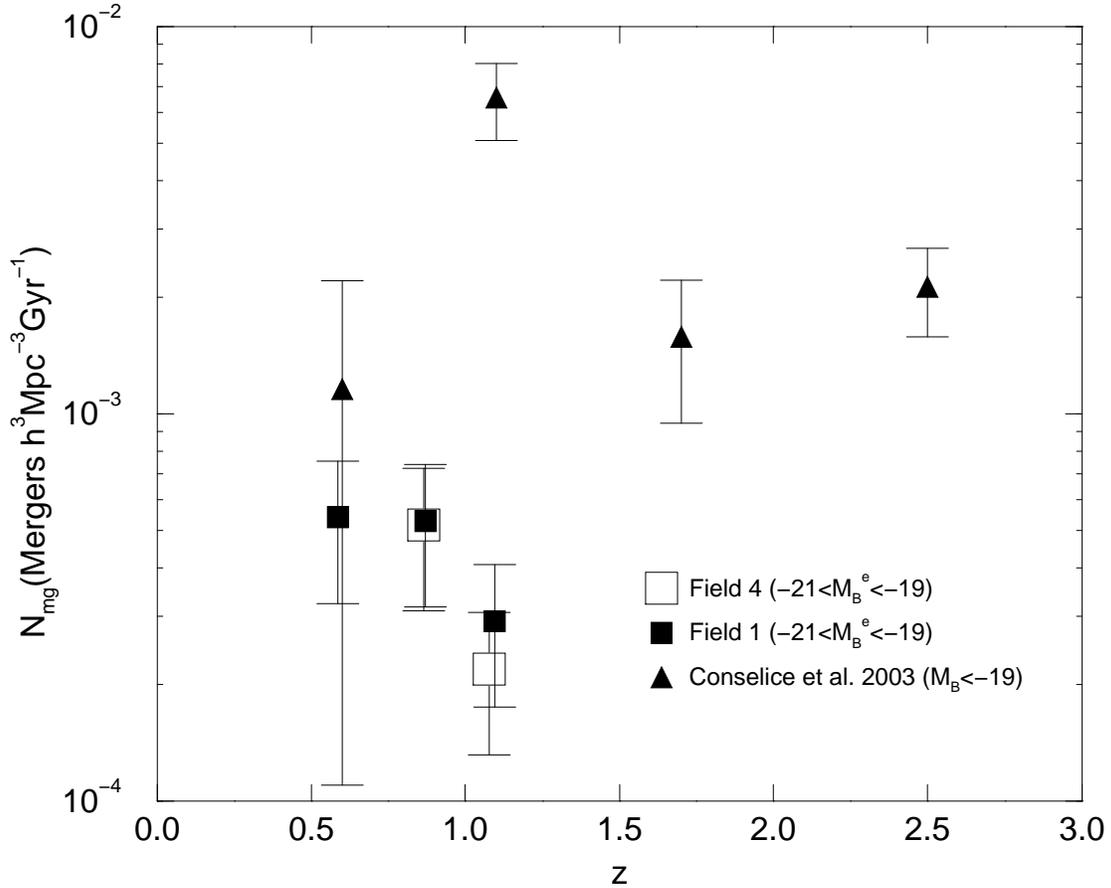} \caption{Comoving volume merger rate, $N_{mg}$,
as a function of  redshift $z$. Estimates from the DEEP2 fields
using equation (4) are marked with symbols as indicated; the data
from Conselice et al. (2003) are represented by the filled
triangles. \label{fig3}}
\end{figure}


\begin{thebibliography}{}
\bibitem[Abraham(1999)]{abr99} Abraham, R. G. 1999, IAU Symp. 186, Galaxy Interactions at Low and High Redshifts (astro-ph/9802033)
\bibitem[Bundy et al.(2004)]{bun04} Bundy, K., Fukugita, M.,Ellis, R. S., Kodama, T., \& Conselice, C. J. 2004, \apj, 601, L123
\bibitem[Bell(2004)]{bel04} Bell, E. F. 2004, preprint (astro-ph/0408023)
\bibitem[Burkey, Keel, \& Windhorst(1994)]{bur94} Burkey, J. M., Keel, W. C., \& Windhorst, R. A. 1994, \apj, 429, L13
\bibitem[Carlberg et al.(1994)]{car94} Carlberg, R. G., Pritchet, C. J., \& Infante, L. 1994, \apj, 435, 540
\bibitem[Carlberg et al.(2000)]{car00} Carlberg, R. G., et al. 2000, \apj, 532, L1
\bibitem[Coil et al.(2004)]{coi04} Coil, A. L., et al. 2004, ApJ, submitted (astro-ph/0403423)
\bibitem[Conselice et al.(2003)]{con03} Conselice, C. J., Bershady, M. A., Dickinson, M., \& Papovich, C. 2003, \aj, 126, 1183
\bibitem[Davis et al.(2003)]{dav03} Davis, M., et al. 2003, SPIE, 4834, 161 (astro-ph/0209419)
\bibitem[Davis, Gerke, \& Newman(2004)]{dav04} Davis, M., Gerke, B., \& Newman, J. A. 2004, preprint(astro-ph/0408344)
\bibitem[Driver \& De Propris(2003)]{dri03} Driver, S., \& DePropris, R. 2003, Ap\&SS, 285, 175
\bibitem[Dubinski, Mihos, \& Hernquist(1996)]{dub96} Dubinski, J., Mihos, J. C., \& Hernquist, L. 1996, \apj, 462, 576
\bibitem[Faber et al.(2003)]{fab03} Faber, S. M., et al. 2003, SPIE, 4841, 1657
\bibitem[Gottl\"{o}ber, Klypin, \& Kravtsov(2001)]{got01} Gottl\"{o}ber, S., Klypin, A., \& Kravtsov, A. V. 2001, \apj, 546, 223
\bibitem[Governato et al.(1999)]{gov99} Governato, F., Gardner, J. P., Stadel, J., Quinn, T., \& Lake, G. 1999, \aj, 117, 1651
\bibitem[Le F\`{e}vre et al.(2000)]{le00} Le F\`{e}vre, O., et al. 2000, \mnras, 311, 565
\bibitem[Mihos \& Hernquist(1994)]{mih94} Mihos, J. C., \& Hernquist, L. 1994, \apj, 425, L13
\bibitem[Mihos (1995)]{mih95} Mihos, J. C. 1995, \apj, 438, L75
\bibitem[Mihos \& Hernquist(1996)]{mih96} Mihos, J. C., \& Hernquist, L. 1996, \apj, 464, 641
\bibitem[Neuschaefer et al.(1997)]{neu97} Neuschaefer, L. W., Im, M., Ratnatunga, K. U., Griffiths., R. E., \& Casertano, S. 1997, \apj, 480, 59
\bibitem[Patton et al.(1997)]{pat97} Patton, D. R., Pritchet., C. J., Yee, H. K. C.,  Ellingson, E., \& Carlberg, R. G. 1997, \apj, 475, 29
\bibitem[Patton et al.(2000)]{pat00} Patton, D. R., Carlberg, R. G., Marzke, R. O., Pritchet., C. J., da Costa, L. N., \& Pellegrini, P. S. 2000, \apj, 536, 153
\bibitem[Patton et al.(2002)]{pat02} Patton, D. R., et al. 2002, \apj, 565, 208
\bibitem[Woods, Fahlman, \& Richer(1995)]{woo95} Woods, D., Fahlman, G. G., \& Richer, H. B. 1995, \apj, 454, 32
\bibitem[Yan, White, \& Coil(2004)]{yan04} Yan, R., White, M., Coil, A. L. 2004, \apj, 607, 739
\bibitem[Yee \& Ellingson(1995)]{yee95} Yee, H. K. C., \& Ellingson, E. 1995, \apj, 445, 37
\bibitem[Yee, Ellingson, \& Carlberg(1996)]{yee96} Yee, H. K. C., Ellingson, E., \& Carlberg, R. G. 1996, \apjs, 102, 269
\bibitem[Zepf \& Koo(1989)]{zep89} Zepf, S. E., \& Koo, D. C. 1989, \apj, 337, 34
\end{thebibliography}
\end{document}